\documentclass[final,5p,times,twocolumn]{elsarticle}
\biboptions{square,sort&compress,comma,numbers}
\pdfoutput=1

\usepackage[latin1]{inputenc}
\usepackage[english]{babel}
\usepackage{amssymb,amsmath,amsthm,cancel,graphicx,xcolor}
\usepackage{booktabs}
\usepackage{mathrsfs}

\usepackage[version=4]{mhchem}

\let\originalleft\left
\let\originalright\right
\renewcommand{\left}{\mathopen{}\mathclose\bgroup\originalleft}
\renewcommand{\right}{\aftergroup\egroup\originalright}

\def\beq{\begin{equation}}  
\def\eeq{\end{equation}}
\def\({\left(}
\def\){\right)}
\def\[{\left[}
\def\]{\right]}

\def\eq#1{{Eq.~(\ref{#1})}}

\def\fig#1{{Fig.~\ref{#1}}}

\def\sect#1{{Sect.~\ref{#1}}}

\journal{Physics Letters B}

\makeatletter
\def\ps@pprintTitle{%
 \let\@oddhead\@empty
 \let\@evenhead\@empty
 \def\@oddfoot{\reset@font\hfil\thepage\hfil}
 \let\@evenfoot\@oddfoot
}
\makeatother

\begin{document}

\begin{frontmatter}



\title{Alpha radioactivity deep-underground as a probe of axion dark matter}


\author[label1]{Carlo Broggini} 
\author[label2]{Giuseppe Di Carlo} 
\author[label1]{Luca Di Luzio} 
\author[label3,label1]{Claudio Toni} 
\address[label1]{Istituto Nazionale di Fisica Nucleare, Sezione di Padova, Via F.~Marzolo 8, 35131 Padova, Italy}
\address[label2]{Istituto Nazionale di Fisica Nucleare, Laboratori Nazionali del Gran Sasso, 67100 Assergi (AQ), Italy}
\address[label3]{Dipartimento di Fisica e Astronomia `G.~Galilei', Universit\`a di Padova, \\ Via F.~Marzolo 8, 35131 Padova, Italy}

\begin{abstract}
We propose to investigate the time modulation of radioisotope decays deep underground as a method to explore axion dark matter. In this work, we focus on the $\alpha$-decay of heavy isotopes and develop a theoretical description for the $\theta$-dependence of $\alpha$-decay half-lives, which enables us to predict the time variation of $\alpha$-radioactivity in response to an oscillating axion dark matter background. To probe this scenario, we have recently constructed and installed a setup deep underground at the Gran Sasso Laboratory, based on the $\alpha$-decay of Americium-241. This prototype experiment, named RadioAxion-$\alpha$, will allow us to explore a broad range of oscillation's periods, from a micro-second up to one year, thus providing competitive limits on the axion decay constant across 13 orders of magnitude in the axion mass, ranging from $10^{-9}$ eV to $10^{-20}$ eV after one month of data collection, and down to $10^{-22}$ eV after three years. 

\end{abstract}


\end{frontmatter}


\section{Introduction}
\label{intro}

By addressing the strong CP problem \cite{Peccei:1977hh,Peccei:1977ur,Weinberg:1977ma,Wilczek:1977pj} 
and the dark matter puzzle \cite{Dine:1982ah,Abbott:1982af,
Preskill:1982cy}, the 
Quantum Chromodynamics (QCD) axion provides a compelling pathway 
beyond the Standard Model of particle physics. In recent years, there has been a 
flourishing 
of new experimental strategies for axion detection (for 
reviews, see e.g.~\cite{Irastorza:2018dyq,DiLuzio:2020wdo,Sikivie:2020zpn}). 
While traditional approaches to axion searches rely on the model-dependent axion coupling to photons, 
a remarkable prediction of the QCD axion stems from the model-independent axion coupling to gluons. 
By promoting the topological $\theta$-term of QCD (defined in \eq{eq:deftheta}) 
to be a time-varying 
axion 
field, one can test the axion-gluon coupling  
through
the oscillating electric dipole moment (EDM) of the neutron,  
induced by the axion dark matter background \cite{Graham:2013gfa,Budker:2013hfa,Stadnik:2013raa}. 
Alternative approaches to constrain the axion-gluon coupling have been discussed 
e.g.~in Refs.~\cite{Blum:2014vsa,Abel:2017rtm,Hook:2017psm,Roussy:2020ily,Schulthess:2022pbp,Balkin:2022qer,Zhang:2022ewz}. 
 
More recently, the authors of Ref.~\cite{Zhang:2023lem} have proposed to look for the time variation  
of the decay rate of certain radioisotopes, focussing 
on the $\theta$-dependence 
of $\beta$-decay,  
previously developed in \cite{Lee:2020tmi}.  
This enabled them to set bounds on the axion coupling to gluons from Tritium decay, 
based on data taken at the European Commission's Joint Research Centre \cite{Pomme2}.
 
The search for a time dependence of the nuclear decay 
rates
started
at the birth of 
radioactivity science. 
As a matter of fact, Madame Curie,
in her Ph.D.~thesis \cite{MadameCurie}, reports on the experiment she 
conducted
to
determine the radioactivity of Uranium at midday and midnight, finding no
difference between the two determinations.
In recent years, several studies have reported a modulation at the per mille level in the decay constants of various nuclei, typically over periods of one year, but also spanning one month or one day 
(see Ref.~\cite{McDuffie:2020uuv} and references therein).  
Conversely, other researchers have found no evidence of such an effect \cite{Pomme1,Pomme2,Pomme3}. 

To clarify this intricate scenario, 
we 
performed 
a few
$\gamma$-spectroscopy experiments 
at the
underground Gran Sasso Laboratory \cite{Bellotti:2012if,Bellotti:2015toa,Bellotti:2013bka,Bellotti:2018jzd}.
The choice of the underground laboratory is, in our opinion, 
a
key point. 
Specifically, the rock overburden suppresses the muon and
neutron flux by six and three orders of magnitude, respectively.  
This reduction renders 
irrelevant 
the impact of the annual time variation of the cosmic ray flux,  
which has an amplitude of 
a few percent 
\cite{Bellotti:2015toa}.
Eventually, we were able to exclude modulations
of the decay constant of radioisotopes
with
amplitudes larger than a few parts 
per 10$^{5}$ in
$\ce{^{137}Cs}$ \cite{Bellotti:2012if}, $\ce{^{222}Rn}$ \cite{Bellotti:2015toa},
$\ce{^{232}Th}$ \cite{Bellotti:2013bka}, $\ce{^{40}K}$
and $\ce{^{226}Ra}$ \cite{Bellotti:2018jzd}, for periods between a few hours and
one year.
 
In this work, we propose to investigate the time modulation of radioisotope decays 
deep underground 
as a method to test axion dark matter.  
Unlike the approach of Ref.~\cite{Zhang:2023lem}, we 
focus 
on the $\alpha$-decay of heavy isotopes 
and 
employ a setup 
designed to explore 
a much broader range of periods, from a
microsecond
to one year.  
From the theoretical side, we have developed a 
framework
for the $\theta$-dependence of $\alpha$-decay half-lives,  
allowing us to predict the time variation of $\alpha$-radioactivity in response to an oscillating axion dark matter background.
On the experimental front, we have constructed and installed a 
prototype setup (RadioAxion-$\alpha$)
at the Gran Sasso Laboratory, based on the $\alpha$-decay of Americium-241. 
The choice of $\ce{^{241}Am}$ is motivated by several factors. 
This isotope has a relatively long half-life of about
432.2 yr 
(approximately stable on the timescale of the measurement)
and it predominantly decays by $\alpha$-emission, 
with a $\gamma$-ray byproduct, 
$\ce{^{241}Am} \to 
\ce{^{237}Np} + \alpha + \gamma(59.5 \, \text{keV})$. 
The resulting $\gamma$-ray can be efficiently detected, using 
for instance a NaI crystal. 
Moreover, $\ce{^{241}Am}$
has been produced in nuclear reactors for decades
and is easily accessible, often used in ionization-type smoke detectors.

In the following, we present the 
theoretical framework for the $\theta$-dependence of $\alpha$-decay and 
describe the experimental setup that we have installed at the Gran Sasso Laboratory.  
We conclude with a sensitivity estimate of RadioAxion-$\alpha$ 
on the axion parameter space and discuss future prospects.

\section{Microscopic theory of $\alpha$-decay}

We consider a 
theory of $\alpha$-decay 
of a heavy isotope, 
$\ce{^{A}_{Z}X} \to \ce{^{A-4}_{Z-2}X} + \alpha$, 
obtained by computing the tunnelling probability 
of the $\alpha$-particle
within a 
WKB framework that employs a microscopic $\alpha$--daughter-nucleus potential \cite{Chowdhury:2005nd,Samanta:2007rg,Sayahi:2019add}.
In the semi-classical approximation, 
the half-life is calculated as \cite{Gamow:1928zz,Gurney:1929zz}\footnote{See Ref.~\cite{Holstein:1996bxl} for a critical assessment of the WKB formula.}
\beq
\label{eq:T12alpha}
T_{1/2}=\frac{\ln2}{\nu_0} \text{exp}(K) \, , 
\eeq
where, in natural units,
\beq
\label{eq:defK}
K=2\int_{r_1}^{r_2} dr \ \sqrt{2\mu[V_{\text{tot}}(r)-Q_\alpha]}
\eeq
is the WKB integral, with $\mu=M_\alpha M_d/(M_\alpha+M_d)\approx M_\alpha$ the reduced mass of the 
$\alpha$--daughter-nucleus
system, 
$Q_\alpha = M(A,Z) - M_d - M_\alpha$
is the energy of the emitted $\alpha$-particle and $r_{1,2}$ the turning points of the potential, 
defined by the conditions
$V_{\text{tot}}(r_1)=V_{\text{tot}}(r_2)=Q_\alpha$.
In \eq{eq:T12alpha}, 
$\nu_0$ denotes the assault frequency, i.e.~the frequency at which the $\alpha$-particle collides against the wall of the potential, which is given by (see e.g.~\cite{Sayahi:2019add})
\beq 
\label{eq:defnu0}
\nu_0 = \frac{1}{2\mu} \[ \int_{0}^{r_1} \frac{dr}{\sqrt{2\mu |Q_\alpha - V_{\text{tot}}(r) | }} \]^{-1} \, . 
\eeq
In the limit of a square potential 
well of depth $-V_0$ this reads $\nu_0 = v / (2 r_1)$, 
with 
$v = \sqrt{2(Q_\alpha + V_0)/\mu}$, 
which can be interpreted as the frequency at which the $\alpha$-particle strikes the barrier 
\cite{Holstein:1996bxl}. 

The central potential among the $\alpha$-particle and daughter nucleus is the sum of the nuclear potential, 
the Coulomb potential and the rotational term, i.e.
\beq
\label{eq:Vtot}
V_{\text{tot}}(\vec{R})=V_N(\vec{R})+V_C(\vec{R})+\frac{\ell(\ell+1)}{2\mu R^2} \, , 
\eeq
where $\ell$ denotes the angular momentum of the nuclear transition and $R = |\vec{R}|$. 
The nuclear potential is obtained by double-folding the densities of the $\alpha$ and daughter nucleus \cite{Satchler:1979ni}
\begin{align} 
\label{eq:VN}
V_N(\vec{R}) &= \int \int \, d^3r_\alpha d^3r_d \, \rho_\alpha (\vec{r}_\alpha) \rho_d (\vec{r}_d) \nonumber \\ 
&\times \tilde{v}(\vec{r} = \vec{r}_d - \vec{r}_\alpha + \vec{R},\rho_\alpha (\vec{r}_\alpha),\rho_d (\vec{r}_d) ) \, , 
\end{align}
where
$\tilde{v}(\vec{r},\rho_\alpha,\rho_d)=v(\vec{r}) \, g(\rho_\alpha,\rho_d)$
is the single-nucleon effective potential with a density-dependent correction.
Since for the $\alpha$-decay process only the iso-scalar component 
of the potential 
contributes,  
we take as an input the iso-scalar term of the so-called M3Y effective potential, 
supplemented by a zero-range potential for the single-nucleon exchange 
\cite{Bertsch:1977sg,Satchler:1979ni,Kobos:1984zz,Gils:1987kay} 
\begin{align} 
\label{eq:M3Y_Reid} 
v_{\text{M3Y}}(\vec{r}) &= \Big[ - 2134 \frac{\exp(-2.5r)}{2.5r} + 7999 \frac{\exp(-4r)}{4r} \nonumber \\
& -276 \, \delta(\vec{r}) \Big] \ \text{MeV} \, ,
\end{align}
with $r = |\vec{r}|$ in units of $1 \, \text{fm} \approx 1 / (198 \, \text{MeV})$. 
Different choices of the nuclear potential, 
such as the so-called Paris version of the M3Y potential \cite{Anantaraman:1983nxh}, 
have a minor impact,  
with differences at the per mille level in the final result of \eq{eq:Itpred}.  
For the density-dependent term, we consider \cite{Chowdhury:2005nd}
\beq
g(\rho_\alpha,\rho_d)=(1-\beta\rho_\alpha^{2/3})(1-\beta\rho_d^{2/3}) \, ,
\eeq
with $\beta=1.6 \, \text{fm}^{2}$. 
Following Ref.~\cite{Satchler:1979ni}, the density distribution for the $\alpha$ particle has been taken to have the Gaussian form
\beq
\rho_{\alpha}(\vec{r})=0.4229 \,\text{exp}(-0.7024r^2) \, \text{fm}^{-3} \, ,
\eeq
whose volume integral is equal to its mass number $A_\alpha=4$. The matter distribution of the daughter nucleus can be instead 
described by a spherically symmetric Fermi function \cite{Chowdhury:2005nd}
\beq
\label{eq:rhoda}
\rho(\vec{r})=\frac{\rho_0}{1+\text{exp}\left(\frac{r-c_d}{a}\right)} \, , 
\eeq
with
$c_d=r_d(1-\pi^2a^2/(3r_d^2))$, $r_d=1.13 \, A_d^{1/3} \, \text{fm}$, $a=0.54 \, \text{fm}$,
while $\rho_0$ is a normalization constant, taken so that the volume integral is equal to the mass number of the daughter particle, 
$A_d = A-4$. 

Finally, the Coulomb potential is given by 
\beq
V_C(\vec{R})=\begin{cases}
\frac{Z_\alpha Z_d \alpha_\text{QED}}{2R_c}\left[ 3-\left( \frac{R}{R_c} \right)^2 \right] \quad \text{for $R<R_c$} \, , \\
\\
\frac{Z_\alpha Z_d \alpha_\text{QED}}{R} \quad \text{for $R>R_c$} \, ,
\end{cases}
\eeq
where $R_c=c_\alpha+c_d$, with $c_\alpha=r_\alpha(1-\pi^2a^2/(3r_\alpha^2))$ and $r_\alpha=1.13 \, A_\alpha^{1/3} \, \text{fm}$.

Within such a framework we are able to reproduce the $\alpha$-decay half-lives
of 
heavy isotopes at the order of magnitude level, in accordance with the results of Refs.~\cite{Chowdhury:2005nd,Samanta:2007rg,Sayahi:2019add}. 
Note that the $\alpha$-decay process is exponentially sensitive to the 
WKB integral, so that the lifetimes span several orders of magnitudes when varying the $Q_\alpha$ value 
for different nuclei.

\section{$\theta$-dependence of $\alpha$-decay} 
\label{sec:thetadep}

The $\theta$-term of QCD is defined by the operator  
\beq 
\label{eq:deftheta}
\mathcal{L}_\theta = \frac{g_s^2 \theta}{32\pi^2} G^a_{\mu\nu} \tilde G^{a\,\mu\nu} \, , 
\eeq
where $|\theta|\lesssim 10^{-10}$ from the non-observation of the 
neutron EDM 
\cite{Abel:2020pzs}. 
The smallness of $\theta$ constitutes the so-called strong CP problem, 
which can be solved by promoting the 
$\theta$-term to be a dynamical field, $\theta \to a(x)/f_a$, 
where $a(x)$ is the axion and $f_a$ a mass scale known as the axion decay constant. 
The axion field acquires a potential in the background 
of QCD instantons and relaxes dynamically to zero, 
thus explaining the absence of CP violation in strong interactions \cite{Peccei:1977hh,Peccei:1977ur,Weinberg:1977ma,Wilczek:1977pj}. 

In the following, we will be interested in the $\theta$-dependence of nuclear quantities, 
anticipating the fact that we will interpret $\theta(t)$ as a time-varying background axion field, 
related to 
the dark matter of the universe \cite{Dine:1982ah,Abbott:1982af,
Preskill:1982cy}.
The consequences of a non-zero $\theta$ in nuclear physics have been previously investigated in 
Refs.~\cite{Ubaldi:2008nf,Lee:2020tmi}, also in connection with the idea of establishing an anthropic bound on $\theta$. 

There are various ways in which the $\theta$-dependence can manifest in nuclear physics, the most prominent is 
through the pion mass \cite{Leutwyler:1992yt,Brower:2003yx}
\beq 
M^2_{\pi}(\theta) = M^2_{\pi} \cos{\frac{\theta}{2}} \sqrt{1 + \varepsilon^2 \tan^2\frac{\theta}{2}} \, , 
\eeq
with $M_{\pi} = 139.57$ MeV and $\varepsilon = (m_d-m_u)/(m_d+m_u)$. The $\theta$-dependence of other 
low-lying resonances, including $\sigma(550)$, $\rho(770)$ and $\omega(782)$ --  
which, along with the pion, are  
responsible for the mediation of nuclear forces in the one-boson-exchange (OBE) 
approximation --  
has been determined based on $\pi\pi$ scattering data
in Ref.~\cite{Acharya:2015pya}.  

A key role 
for the binding energy 
of heavy nuclei 
is played by the $\sigma$ and $\omega$ channels, 
via the contact interactions \cite{Furnstahl:1999rm}
\beq 
H = G_S (\bar N N) (\bar N N) + G_V (\bar N \gamma_\mu N) (\bar N \gamma^\mu N) \, , 
\eeq
which control, respectively, the scalar (attractive) and vector (repulsive) part of the nucleon-nucleon interaction 
\cite{Donoghue:2006rg,Damour:2007uv}. 
To describe their $\theta$-dependence we employ the following parametrization
\beq 
\eta_S = \frac{G_S(\theta)}{G_S(\theta=0)} \, , \quad 
\eta_V = \frac{G_V(\theta)}{G_V(\theta=0)} \, .
\eeq
In Ref.~\cite{Donoghue:2006du} it was found that 
the pion mass dependence of $\omega$ exchange 
leads to subleading corrections compared to the effects related to the 
$M^2_{\pi}$ sensitivity of the scalar channel. Hence, to a good approximation, we can take $\eta_V = 1$ 
and consider only the leading $\theta$-dependence in the scalar channel, 
which is described by the following fit \cite{Ubaldi:2008nf} to 
Fig.~2 in \cite{Damour:2007uv} 
\beq
\label{eq:etaStheta}
\eta_S (\theta) = 1.4 - 0.4 \frac{M^2_{\pi}(\theta)}{M^2_{\pi}} \, .
\eeq
Moreover, based on the relativistic mean-field simulations of \cite{Furnstahl:1999rm} 
for two specific nuclei, Ref.~\cite{Damour:2007uv} 
finds that the variation of the binding energy (BE) for a nucleus of mass number $A$
can be written as (keeping only the variation due to $\eta_S(\theta)$) 
\begin{align}
\label{eq:BEtheta}
\text{BE}(\theta)&=\text{BE}(\theta=0) +(120 A - 97 A^{2/3})(\eta_S (\theta) -1) \, \text{MeV} \, , 
\end{align}
where the terms proportional to $A$ and $A^{2/3}$ represent a volume and surface contribution, 
in analogy to the semi-empirical mass formula \cite{Weizsacker:1935bkz}. 
Note that \eq{eq:BEtheta} can be safely used only in the small $\theta$-value regime, 
since the underlying nuclear model relies on a strong cancellation between 
the attractive $\sigma$ channel and the repulsive $\omega$ channel. 

Hence, substituting the expressions of the BEs above in the definition of 
$Q_\alpha =  \text{BE}(A-4,Z-2) + \text{BE}(4,2) - \text{BE}(A,Z)$, 
we find 
\begin{align}
\label{eq:Qalphatheta}
Q_\alpha(\theta)&=Q_\alpha(\theta=0) -97 \text{ MeV }(\eta_S (\theta) -1) \nonumber \\
&\times ((A-4)^{2/3}+4^{2/3}-A^{2/3}) \, .
\end{align}
It turns out that $Q_\alpha(\theta)$ provides, by far, the leading effect in order to assess the $\theta$-dependence 
of $\alpha$-decay.\footnote{A similar observation was noted in the context of the variation of the fine-structure constant, 
which impacts primordial nuclear abundances \cite{Meissner:2023voo}.} 
Other possible dependences from $\theta$ arise  
from the reduced mass $\mu$, see e.g.~\eq{eq:defK},   
and the M3Y nuclear potential. 
The former is given by $\mu(\theta) \approx 4 m_N(\theta) - \text{BE}_4(\theta)$, 
where $m_N(\theta)$ was computed in Ref.~\cite{Lee:2020tmi}, while $\text{BE}_4(\theta)$ 
can be read from \eq{eq:BEtheta} with $A=4$.  
Regarding instead the $\theta$-dependence of the M3Y potential, this
can be implemented by interpreting the exponential terms in 
\eq{eq:M3Y_Reid} as arising from $\sigma$ (attractive) and $\omega$ (repulsive) exchange in the OBE
approximation. Focussing on the leading 
$\theta$-dependence 
from $\sigma$ exchange, 
we can rescale the pre-exponential and exponential factors respectively 
via $g^2_{\sigma NN} (\theta)$ and $M^2_{\sigma} (\theta)$.  
The $\theta$-dependence of the $\sigma$ mass is taken from \cite{Acharya:2015pya,Lee:2020tmi}, 
while the $\sigma$ coupling can be expressed in terms of 
$G_S (\theta) = - g^2_{\sigma NN} (\theta) / M^2_\sigma (\theta)$.  
We find that the $\theta$-dependence arising 
from both the reduced mass and 
the nuclear potential 
remains always subleading with respect to that from $Q_\alpha(\theta)$, 
basically below the percent level in the final result of \eq{eq:Itpred}.  
The predominance of the $\theta$-dependence 
of $Q_\alpha(\theta)$ with respect to that of $V_{\rm tot}(\theta)$ in \eq{eq:defK} can be understood 
by the fact that 
the WKB integral is defined across the potential barrier, and the latter 
is dominated by the Coulomb potential that is not affected 
by $\theta$.

\section{Axion dark matter time modulation}

Assuming an oscillating axion dark matter field from misalignment \cite{Dine:1982ah,Abbott:1982af,
Preskill:1982cy}, 
the time dependence of the $\theta$ angle can be approximated as
$\theta(t)=\theta_0 \cos(m_a t)$, 
with
\begin{align}
\label{eq:theta0DM}
\theta_0&=\frac{\sqrt{2\rho_{\rm DM}}}{m_a f_a} \, , 
\end{align}
in terms of $\rho_{\rm DM} \approx 0.45 \, \text{GeV/cm}^3$.
For
a standard QCD axion, one has 
\beq 
\label{eq:QCDband}
m_a f_a = \frac{\sqrt{m_u m_d}}{m_u + m_d} m_\pi f_\pi = (76 \ \text{MeV})^2 \, , 
\eeq
corresponding to $\theta_0 = 5.5 \times 10^{-19}$. In the following, we will treat $m_a$ and $f_a$ as independent parameters and discuss the sensitivity 
of $\alpha$-decay observables in the $(m_a, 1/f_a)$ plane. 

Following Ref.~\cite{Zhang:2023lem}, we introduce the 
observable
\beq 
\label{eq:Itdef}
I(t) \equiv \frac{T^{-1}_{1/2}(\theta(t)) - \langle T^{-1}_{1/2} \rangle }{\langle T^{-1}_{1/2} \rangle } \, , 
\eeq
where $\langle T^{-1}_{1/2} \rangle$ denotes a time average.  
Given that the main 
$\theta$-dependence in \eq{eq:etaStheta} arises through the pion mass, 
we expect that 
$T_{1/2}(\theta)$ is analytic in $\theta^2$ and 
admits the Taylor expansion\footnote{This is also verified a posteriori by a numerical fit of the half-life 
as a function of $\theta$.}
\beq 
\label{eq:Texpansion}
T_{1/2}(\theta) \approx T_{1/2}(0) + \mathring{T}_{1/2}(0) \theta^2 \, , 
\eeq 
where we introduced the derivative symbol, $\mathring{f} \equiv df/d\theta^2$. 
Since $\theta^2 \ll 1$, \eq{eq:Texpansion} does provide an excellent approximation to the full 
$\theta$-dependence, which is anyway taken into account in our numerical analysis. 
Using $\langle \cos^2 (m_a t) \rangle = 1/2$ 
and expanding at the first non-trivial order in $\theta_0$, 
we find 
\begin{align} 
\label{eq:Itpred}
I(t) &\approx - \frac{1}{2} \frac{\mathring{T}_{1/2}(0)}{T_{1/2}(0)} \theta_0^2 \cos(2 m_a t) \nonumber \\
&= - 4.3 \times 10^{-6} \cos(2 m_a t) \left( \frac{\rho_{\rm DM}}{0.45 \, \text{GeV/cm$^3$} } \right) \nonumber \\
&\times \left( \frac{10^{-16} \, \text{eV}}{m_a} \right)^2  \left( \frac{10^{8}  \, \text{GeV}}{f_a} \right)^2
\, ,  
\end{align}  
where $\mathring{T}_{1/2}(0) / T_{1/2}(0) \approx 125$ has been obtained 
 by fitting the numerical expression 
of $T_{1/2}(\theta)$ at small $\theta$ values. 
To obtain \eq{eq:Itpred} we also used $Q_{\alpha}(\theta=0) = 5.486$ MeV,  
corresponding to the dominant $\alpha$-decay transition 
$\ce{^{241}Am} \to 
\ce{^{237}Np}^\star + \alpha$
with $\ell = 0$, 
and substituted $\theta_0$ from \eq{eq:theta0DM}. 
An analytical approximation for $\mathring{T}_{1/2}(0) / T_{1/2}(0)$ 
valid for the $\alpha$-decay of a generic $\ce{^{A}_{Z}X} $ isotope is provided in \ref{app:thetadependencealpha}. 

Note that the 
large theoretical uncertainty in the prediction of $T_{1/2}(0)$, stemming from its 
exponential dependence from the WKB integral $K$, 
is washed out thanks to the normalization of \eq{eq:Itdef}. 
In fact, neglecting the small $\theta$-dependence arising from $\nu_0$ in \eq{eq:defnu0}, 
amounting to an effect below the per mille level in \eq{eq:Itpred}, 
we have $\mathring{T}_{1/2}(0) / T_{1/2}(0) \approx \mathring{K}(0)$.  

\section{Experimental setup} 

To study the time modulation 
of the $\alpha$-decay of 
$\ce{^{241}Am}$,  
we built
a prototype setup
(RadioAxion-$\alpha$)
which
we installed deep underground at the Gran Sasso Laboratory, in a dedicated
container.
A $3" \times  3"$ NaI crystal 
detects the $\gamma$-rays due to the $\alpha$-decay of $^{241}$Am, 
primarily ($85 \%$ of the time) at 59.5 keV, and the X-rays
from $^{237}$Np atomic transitions.  
The signal from the photomultiplier is processed by an ORTEC digiBASE-E, a
14-pin photomultiplier tube base that is directly connected to the
photomultiplier.
The digiBASE, the photomultiplier, the crystal and the source, kept in a fixed
position in front of the crystal, 
are closed inside a parallelepiped made of polyethylene, completely surrounded by
a passive
shielding of 5 cm of copper and 10 cm of lead,
in order to suppress the laboratory  $\gamma$-ray background.
Data acquisition operates in list mode, i.e.~each signal above the 10 keV
threshold is converted to a digital value which is transmitted to the computer
along with the time of the event. The time resolution is 160 ns.
To mitigate the impact of the digiBASE's quartz aging, 
we also acquire a signal 
every
second, 
generated by an FS725 10 MHz Rb Frequency Standard which has a 20 year
aging factor of less than 5 $\times$ 10$^{-9}$.

\begin{figure}[t!]
\vspace{-0.5cm}
\centerline{
\ \ \ \includegraphics[width=0.58\textwidth,angle=0]{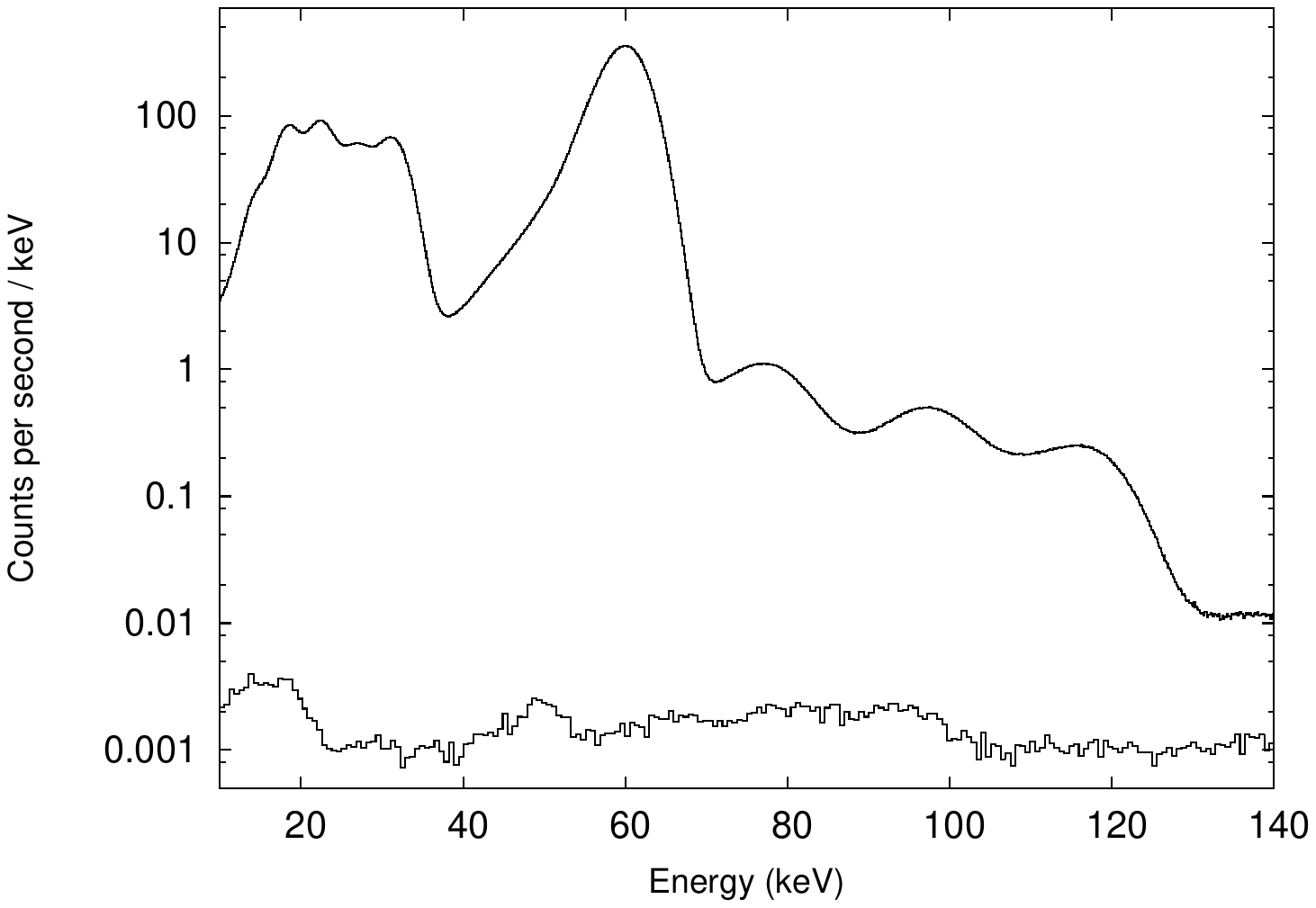}
}
\vskip -5mm
\caption{$\gamma$-spectrum (counts per second per keV) of the  $^{241}$Am source (upper curve) compared 
to the background (lower curve). 
The dominant contribution arises from the $\gamma$ at 59.5 keV.}
\label{fig:Am_fig1}
\end{figure}

In Fig.~\ref{fig:Am_fig1} we show the energy spectrum of the events collected in
24 hours, with and without the $\ce{^{241}Am}$ source. 
With the $\ce{^{241}Am}$ source we have a rate 
(counts per second)
of about 4 kHz, to be compared to a
background of 0.2 Hz. 
The background, i.e.~the counts in the absence of the source, is
essentially due to the inner radioactivity of the NaI crystal, 
while the background due to the 
cosmic ray flux 
is safely negligible. 

\section{Sensitivity estimate} 

The theoretical prediction in \eq{eq:Itpred} 
can be 
compared with 
$I_{\rm exp} (t) \equiv (N(t) - \langle N \rangle) / \langle N \rangle$,  
where $N(t)$ is the observed number of events in a given interval of time and 
$\langle N \rangle$ its expected value, according to the exponential decay 
law.
Potential sources of systematic errors include the detection of $\gamma$-rays and their time-stamping. 
The former is mitigated by operating the NaI detector well-below the radiation damage threshold 
and by the reduced background in the underground environment. 
The latter is handled thanks to the precision of a Rb atomic clock.
Hence, we expect our uncertainties to be statistically dominated in the current setup. 

We started data taking at the beginning of May 2024. 
With a rate of about 4 kHz events, we expect to reach a 2$\sigma$ error 
of $2 / \sqrt{4000 / s \times \pi \times 10^7 s} \approx 6 \times 10^{-6}$ 
on 
$I_{\rm exp}$ after one year of data taking.  
Given the 160 ns time resolution of our setup 
and referring to the oscillation period as
$\Delta t$, 
we consider two realistic benchmarks corresponding to distinct experimental phases:
$i)$ Phase 1: $1 \, \mu\text{s} < \Delta t < 10 \ \text{days}$ and 
$I_{\rm exp} = 2 \times 10^{-5}$ at 2$\sigma$
with one month of data taking and 
$ii)$ Phase 2: $1 \, \mu\text{s} < \Delta t < 1 \, \text{yr}$ and $I_{\rm exp} = 4 \times 10^{-6}$ 
at 2$\sigma$
with three years of data taking. 

The sensitivity of the present experiment is ultimately limited by the number of detected
events due to the $\ce{^{241}Am}$ source. 
By increasing the source activity by a factor of 10,  
it would be possible to improve the sensitivity by a factor of 3. 
Further improvements would require, 
in addition to a more powerful $\ce{^{241}Am}$ source, a 
faster detector, for instance a plastic scintillator, and a 
significantly upgraded
data acquisition system. All in all, an improvement of 
up to two
orders of magnitude in the sensitivity could be possible with a set-up similar to ours 
but with more cutting-edge technologies. 

 \begin{figure}[t!]
 \centering
\!\!\!\!\!\!\!    \includegraphics[width=0.50\textwidth]{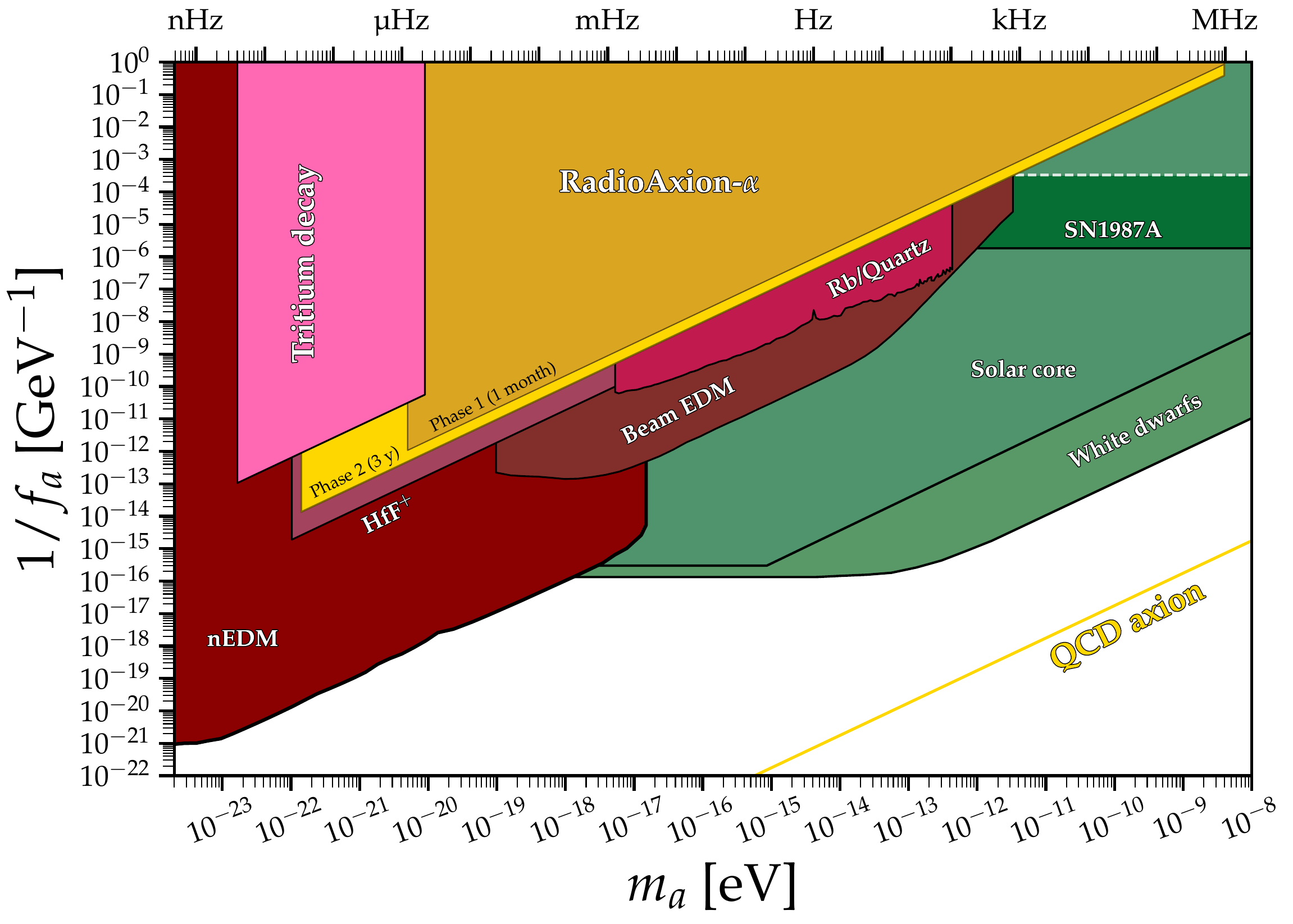}
     \caption{Constraints on the axion dark matter coupling to gluons. 
     The projected sensitivities of RadioAxion-$\alpha$ 
     are displayed for two experimental phases (yellow-shaded areas). 
     Limits from laboratory experiments (red-shaded areas) 
     and astrophysics (green) are shown as well for comparison.
     Figure adapted from \cite{AxionLimits}.}
     \label{fig:sensitivity}
 \end{figure}

The results of our sensitivity estimate 
for the two experimental phases of RadioAxion-$\alpha$
are shown in \fig{fig:sensitivity}. 
For comparison, we also display laboratory limits from 
EDM searches \cite{Abel:2017rtm,Roussy:2020ily,Schulthess:2022pbp}, 
radio-frequency atomic transitions \cite{Zhang:2022ewz}, 
and Tritium decay \cite{Zhang:2023lem}, 
as well the SN 1987A bound stemming from the 
axion-nucleon EDM coupling \cite{Graham:2013gfa,Lucente:2022vuo} 
and finite-density-induced bounds from the solar core and white dwarfs \cite{Hook:2017psm,Balkin:2022qer}. 
Note that for 
$1/f_a \gtrsim 3.3 \times 10^{-4} \, \text{GeV}^{-1}$ 
(above the dashed line in \fig{fig:sensitivity})
axions enter the trapping regime and the 
cooling bound from SN 1987A does not apply \cite{Lucente:2022vuo}. 

The yellow, QCD axion line stems from the relationship in \eq{eq:QCDband},   
but it remains beyond the reach of the techniques proposed here.
The standard $m_a$--$f_a$ relation can however be modified
in such a way that the axion mass is suppressed for fixed $f_a$ 
through a symmetry principle
\cite{Hook:2018jle,DiLuzio:2021pxd,DiLuzio:2021gos,Banerjee:2022wzk}.  
This can be achieved by employing 
$\mathcal{N}$ mirror copies of the Standard Model, 
endowed with a $Z_{\mathcal{N}}$ symmetry, under which 
$\text{SM}_k \to \text{SM}_{k+1(\text{mod}\, \mathcal{N})}$
and the axion 
acting non-linearly: 
$a \to a + 2\pi k / \mathcal{N}$, with $k = 0,\ldots, \mathcal{N}-1$. 
It can be shown \cite{Hook:2018jle,DiLuzio:2021pxd} that 
this results in the axion mass being exponentially suppressed as
$z^{\mathcal{N}/2}$, with $z = m_u/m_d \approx 0.5$, 
compared to the usual axion mass.  
Additionally, a modified version of the misalignment mechanism 
can still support the possibility of axion dark matter \cite{DiLuzio:2021gos}.

\section{Conclusions}

Our investigation into the time modulation of radioisotope decays deep underground at the Gran Sasso Laboratory has successfully established the RadioAxion-$\alpha$ experiment. This setup, centered on the $\alpha$-decay of $\ce{^{241}Am}$, 
will allow us to cover a wide range of oscillation periods from microseconds to a year. 
Based on realistic projected sensitivities, we will provide with just few years of data  
competitive constraints on the axion decay constant, spanning 13 orders of magnitude in axion mass, from 
from $10^{-9}$ eV to $10^{-22}$ eV. 
We anticipate a better sensitivity compared to existing experiments based on radioactivity, 
such as Tritium decay, and moderately weaker than radio-frequency atomic transitions, which are both sensitive to $\theta^2(t)$. 
On the other hand, EDM-like searches still remain the most effective ones, since they 
depend linearly from $\theta(t)$.

This work not only marks an additional step
in axion dark matter research but also lays the groundwork for a broader project aimed at optimizing the study of the 
$\theta$-dependence of radioactivity. Future efforts will focus on identifying the most effective decay types and isotopes to fully leverage the unique underground environment for axion detection.

\section*{Acknowledgments}

We thank Francesco D'Eramo 
and Massimo Pietroni 
for discussions at the initial stages of this project, 
as well as Fabio Zwirner for suggestions on $\alpha$-decay theory  
and Roberto Isocrate for the improvements of the electronics.
The work of LDL and CT is supported
by the European Union -- Next Generation EU and
by the Italian Ministry of University and Research (MUR) 
via the PRIN 2022 project n.~2022K4B58X -- AxionOrigins.  
LDL is also supported by the European Union -- NextGeneration 
EU and by the University of Padua under the 2021 
STARS Grants@Unipd programme 
(Acronym and title of the project: CPV-Axion -- Discovering the CP-violating axion). 

\appendix
\section{Analytical framework for the $\theta$-dependence of $\alpha$-decay}
\label{app:thetadependencealpha}

In this Appendix we provide an analytical approximation for the 
$\theta$-dependence of $\alpha$-decay, parametrized via the factor 
$\mathring{T}_{1/2} (0) / T_{1/2} (0)$ in \eq{eq:Itpred}, that applies to 
nuclei with generic values of $A$ and $Z$. In the limit of a squared potential well of depth $-V_0$, the total potential reads (assuming $\ell=0$)
\beq
V_{\text{tot}}(\vec{R})=\begin{cases}
-V_0 \quad \text{for $R<R_{\text{well}}$} \, , \\
\\
\frac{Z_\alpha Z_d \alpha_\text{QED}}{R} \quad \text{for $R>R_{\text{well}}$} \, ,
\end{cases}
\eeq
where $R_{\text{well}}$ is the radius of the well, approximately given by $R_{\text{well}}\approx R_0 \ A^{1/3}$, 
with $R_0 \approx 1.13$ fm. 
Such expression yields the following analytical result for the WKB integral
\beq
\label{eq:Zana}
K=Z_\alpha Z_d \alpha_\text{QED}\left(\frac{8\mu}{Q_\alpha}\right)^{1/2}F\(\frac{Q_\alpha R_{\text{well}}}{Z_\alpha Z_d \alpha_\text{QED}}\) \ ,
\eeq
with
\beq
\label{eq:Ffunc}
F(x)=\arccos\sqrt{x}-\sqrt{x}\sqrt{(1-x)}\approx\frac{\pi}{2}-2\sqrt{x}+\dots \, , 
\eeq
where in the last step we have considered the $x \ll 1$ regime, 
that is typically realized for $\alpha$-decay. 
Note that 
the factor $V_0$ drops out from 
$K$ in \eq{eq:Zana},
while $R_{\text{well}}$ enters only through $F(x)$, 
which is a constant in the $x \to 0$ limit. 
Therefore, as anticipated at the end of \sect{sec:thetadep},  
$Q_\alpha$ provides the leading contribution 
compared e.g.~to the nuclear potential.    
Thus one gets
\beq
\frac{\mathring{T}_{1/2}(0)}{T_{1/2}(0)} \approx \mathring{K}\approx\frac{\partial K}{\partial Q_\alpha} \ \mathring{Q}_\alpha
\ ,
\eeq
with
\begin{align}
\mathring{Q}_\alpha & \approx 4.23 \ [ A^{2/3}-(A-4)^{2/3}-4^{2/3} ] \ \text{MeV} \nonumber \\
&\approx -4.23 \ \text{MeV} \ \left[4^{2/3}-\frac{8}{3A^{1/3}}\right] \, ,
\end{align}
from Eq.~\eqref{eq:Qalphatheta}.
Expanding $F(x)$ as in \eq{eq:Ffunc}, our final estimate gives
\beq
\label{eq:TdotsuTgeneral}
\frac{\mathring{T}_{1/2}(0)}{T_{1/2}(0)}\approx 8.45 \ (Z-2) \ \left[4^{2/3}-\frac{8}{3A^{1/3}}\right] \ \left(\frac{\text{MeV}}{Q_\alpha}\right)^{3/2} \ .
\eeq
For instance, for the case of the $\alpha$-decay of $\ce{^{241}Am}$, \eq{eq:TdotsuTgeneral} yields $\mathring{T}_{1/2} (0) / T_{1/2} (0) \approx 128$, 
while keeping the full $F(x)$ dependence in $K$ one gets 125, in excellent agreement with the numerical result in \eq{eq:Itpred}. 

\begin{figure}[t!]
\vspace{-0.5cm}
\centerline{
\includegraphics[width=0.5\textwidth,angle=0]{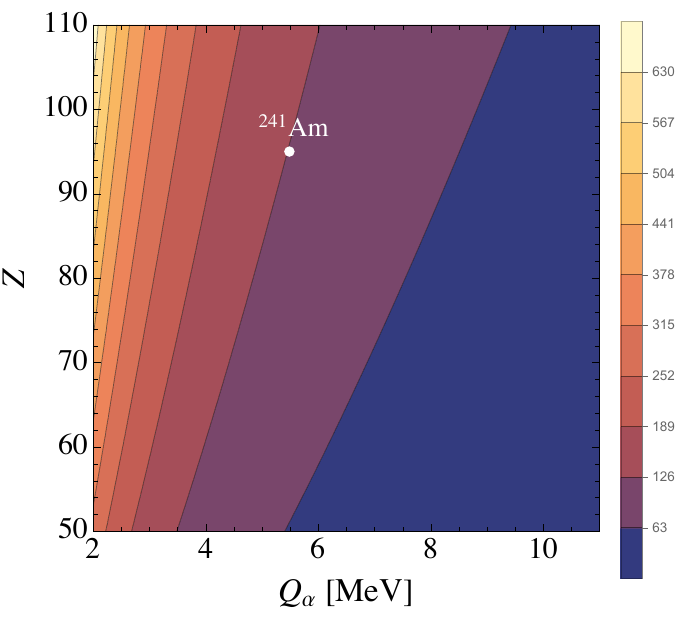} 
}
\caption{Contours of $\mathring{T}_{1/2} (0) / T_{1/2} (0)$ in the $(Q_\alpha, Z)$ plane for $A=241$. 
The case of $\ce{^{241}Am}$ is indicated by a white dot.}
\label{fig:alphalandscape}
\end{figure}

\eq{eq:Ffunc} also shows that the main $\theta$-dependence arises through $Q_\alpha$ and $Z$, with a 
weaker dependence from $A$. In Fig.~\ref{fig:alphalandscape} we display the contour values of $\mathring{T}_{1/2} (0) / T_{1/2} (0)$ 
in the $(Q_\alpha, Z)$ plane, keeping the full $F(x)$ dependence in $K$ and setting $A=241$. In fact, for values $A\gtrsim 100$
relevant for $\alpha$-decays, the $A$ dependence turns out to be rather weak. Hence, we conclude that $\ce{^{241}Am}$ 
performs rather well in the landscape of possibilities for probing the $\theta$-dependence of $\alpha$-decay.

\bibliographystyle{elsarticle-num} 
\bibliography{bibliography}

\end{document}